\documentclass[runningheads]{llncs}
\usepackage{graphicx}
\usepackage{amsmath}
\usepackage{amssymb}
\usepackage{algorithmic}
\usepackage[ruled, vlined]{algorithm2e}
\usepackage{multirow}

\usepackage{url}

\usepackage{breakurl}
\usepackage[breaklinks]{hyperref}
\usepackage{float}
\usepackage{xcolor}
\graphicspath{ {./images/} }

\begin{document}

\title{Voxlines: Streamline Transparency through Voxelization and View-Dependent Line Orders}
\titlerunning{Transparency through Voxelization and View-Dependent Line Orders}

\author{Besm Osman\and
    Mestiez Pereira\and
    Huub van de Wetering\and
    Maxime Chamberland}
\authorrunning{B. Osman et al.}

\institute{Eindhoven University of Technology, Eindhoven, The Netherlands
\email{\{b.osman@student.,m.chamberland@\}tue.nl}
}
\maketitle 
\begin{abstract}
    As tractography datasets continue to grow in size, there is a need for improved visualization methods that can capture structural patterns occurring in large tractography datasets. Transparency \index{transparency} is an increasingly important aspect of finding these patterns in large datasets but is inaccessible to tractography due to performance limitations. In this paper, we propose a rendering method that achieves performant rendering of transparent streamlines, allowing for exploration of deeper brain structures interactively. The method achieves this through a novel approximate order-independent transparency \index{order-independent transparency} method that utilizes voxelization and caching view-dependent line orders per voxel. We compare our transparency method with existing tractography visualization software in terms of performance and the ability to capture deeper structures in the dataset.

\end{abstract}
\section{Introduction}
Tractography datasets are growing larger, posing challenges for their visualization in terms of performance and usability. Various methods have been developed to address performance issues, such as removing colinear points or compressing the dataset\cite{dealing-with-millions}. Existing tractography visualization tools like MRtrix3 \index{MRTrix3} \cite{MRtrix3} and TrackVis \index{TrackVis} \cite{TrackVis} provide filtering options for visualizing different parts of the dataset. Recently, transparency-based methods have emerged to improve usability, such as applying varying transparency to each fiber based on orientation to highlight underlying tissue configurations \cite{orientation-dependent-transparency}. Transparency can play an important role in tractography visualization by enabling exploration of deeper brain structures. Transparency has been used to better convey the spatial relationship between streamlines and surfaces that illustrate their anatomical context. \cite{Schultz_Sauber_Anwander_Theisel_Seidel_2008} However, existing tractography visualization software, utilize subpar transparency methods due to performance constraints \cite{MRtrix3-issue}, limiting the benefits gained from transparency.

\subsection{Existing Transparency Methods}
The basic approach to achieving transparency is by rendering the objects furthest from the screen first, which requires sorting every translucent object whenever the view changes. This method is suitable for visualization contexts with a limited number of transparent objects or a fixed viewing direction. However, this solution is not feasible for interactive tractography visualization. It is not possible to determine a consistent sorting order for entire streamlines, since different parts of a streamline can be at varying distances from the view. Therefore, sorting needs to be done at the level of individual line segments that make up the streamlines. The scale of the tractography dataset makes the computational cost of real-time sorting of line segments impractical. Hence, order-independent methods are required to achieve tractography visualizations with transparency. These techniques eliminate the need for ordering translucent objects. Various order-independent transparency techniques have been researched for 3D line sets, including depth peeling \cite{NVIDIA-Transparency}, multi-layer alpha blending \cite{multi-layer-alpha-blending}, and raycast techniques \cite{voxel-raycast}. In a recent comprehensive study \cite{comparison}, different order-independent transparency techniques for 3D linesets were compared, each showing different advantages and drawbacks. Some techniques require multiple render passes, resulting in decreased performance. Other techniques rely on complex render pipelines supported only by newer hardware or have high preprocessing times.

\section{Method Overview}
We propose a novel method for voxel-based streamline rendering that achieves fast, approximate order-independent transparency without relying on modern render pipelines and minimal preprocessing time. The first part of this method involves splitting the dataset into voxels, where each voxel stores streamline segments that pass through it as described in sections \ref*{voxelization} and \ref*{generating-voxlines}. Next, a mesh \index{mesh} is generated for each voxel that connects the streamlines segments and renders the streamlines explained in Section \ref*{mesh-generation}. This mesh generation is performed once when reading the dataset. During rendering, we sort the voxels from back to front every frame, solving the most noticeable transparency issues. Furthermore, we extend this method by storing a set of precomputed line segment render orders per voxel and selecting the closest ordering based on the current view as described in Section \ref*{view-dependent-line-orders} to improve the transparency accuracy within a voxel.

\subsection{Voxelization} \label{voxelization}
\begin{figure}[h]
    \includegraphics[width=\textwidth]{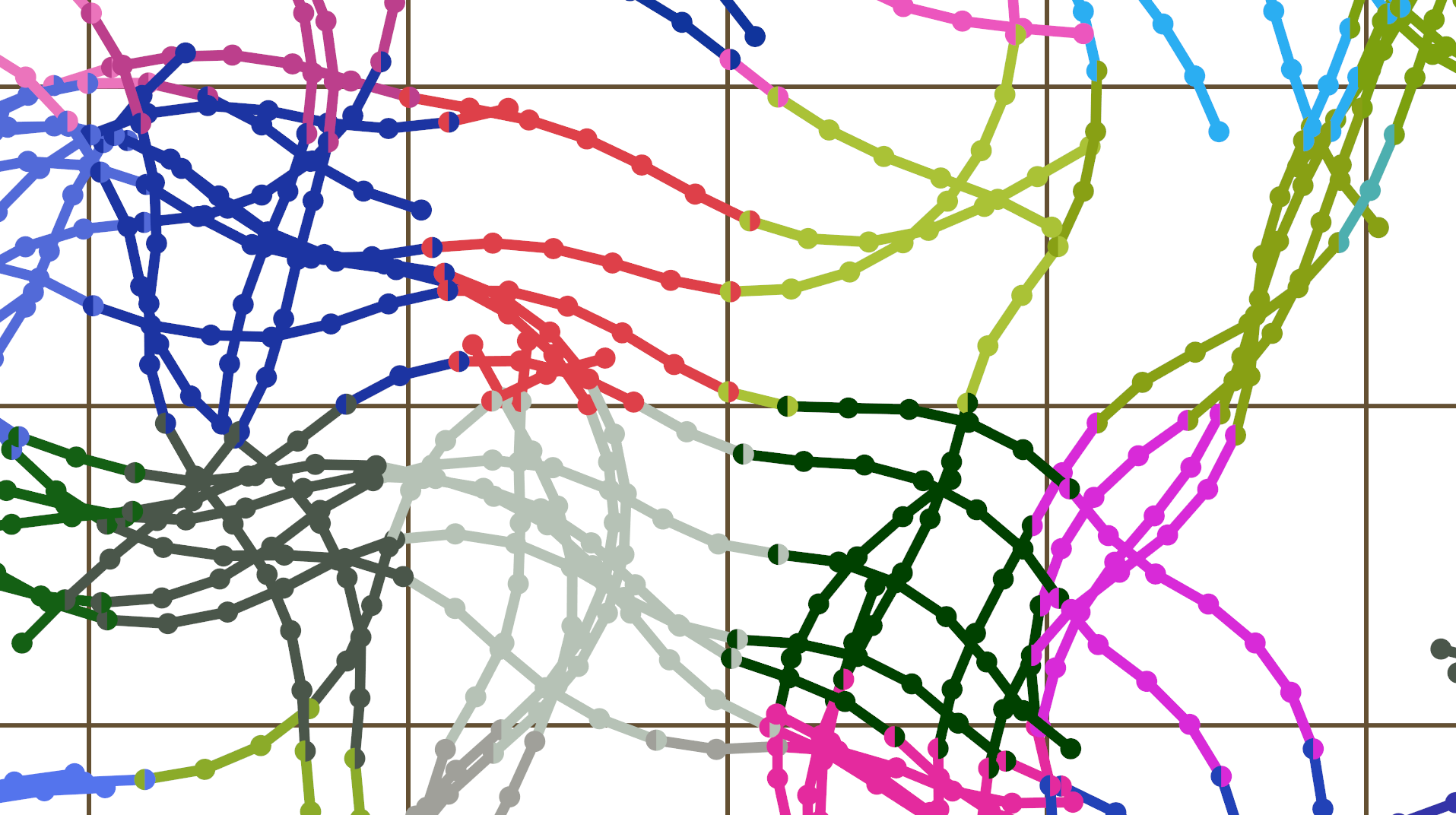}
    \caption{This figure shows a 2D representation of the voxel meshes. Each square in the figure corresponds to a voxel. The points and lines in the figure are colored according to their respective mesh. Some points are used twice to connect voxels, and to illustrate this these points are shown with both colors.} \label{voxel-visualization}
\end{figure}
The algorithm takes a set of streamlines as input, where each streamline consists of a sequence of points. Our goal is to divide the dataset into voxels so that each voxel contains all line segments passing through it. In this paper, we define a voxel as a 3D cube with a fixed position and size. The size of the voxels used in this method differs from voxel sizes used in MRI scans or tractography algorithms. The voxels used in our method are an order of magnitude larger than the voxel sizes used in MRI scans to encompass more streamline points per voxel. To obtain the voxel coordinate of a streamline point $p$, we first calculate its position relative to the dataset by subtracting the minimum bound $\boldsymbol{B_{min}}$, which is defined as the smallest coordinates contained in the dataset: $\boldsymbol{p_{rel}} = \boldsymbol{p} - \boldsymbol{B_{min}}$. This ensures that each voxel has non-negative coordinates. The voxel coordinates $\boldsymbol{v}$ are then obtained by dividing $\boldsymbol{p_{rel}}$ by the voxel size $\boldsymbol{s}$ and taking the floor of the resulting vector: $\boldsymbol{v} = \lfloor \boldsymbol{p_{rel}} / \boldsymbol{s} \rfloor$.

\subsection{Generating Voxlines}\label{generating-voxlines}
Intuitively, grouping points of a streamline based on which voxel the points fall in, can be seen as dividing the streamline points into several parts when a streamline passes through multiple voxels. This can be seen in Figure \ref*{voxel-visualization} where each streamline is split based on the voxel coordinate of each streamline point. In the rest of the paper, we will refer to these parts as "voxlines". A voxline is defined as a sequence of consecutive points of a streamline that is bounded by a voxel. In this context, consecutive means that there are no missing points between the minimum and maximum points of a group of streamline points. For example, points with streamline indices $\{6, 7, 8, 9\}$ would be considered consecutive, while $\{3, 4, 12, 13\}$ would not.

To generate voxlines making up the streamlines we can split each streamline in voxlines based on the voxel each point falls within. However, using only the points in each voxline would result in gaps between voxels when rendering the voxlines. Consider a streamline that is separated into two voxlines. If we render these voxlines as two distinct sets of lines, there will be a gap between them because no line is drawn between the end of the first voxline and the beginning of the second voxline. So defining the voxlines by using only the points that fall within the voxel would cause lines that cross the voxel borders to not be rendered. To address this issue, when generating the voxlines we add an additional point for every voxline that does not contain the final streamline point. This additional vertex corresponds to the next point in the streamline that falls outside of the voxel. By doing this, we fill all the gaps between the voxlines. This is illustrated in Figure \ref{voxel-visualization} by certain voxlines having an extra point outside of their voxel boundary and are thus part of both voxel meshes.

\begin{algorithm}[h] \label{voxlines-algo}
    \DontPrintSemicolon
    \KwData{Input data consist of a sequence $streamlines$ where each element consists of a sequence of points, bound minimum $B_{min}$ of the whole dataset and voxel size $s$ blue{.}}
    $voxelset = \emptyset$ //Empty hashmap with voxel coordinates as keys and sets of voxlines as values\\
    \ForAll{streamlines $S= <p_0, p_1, ... p_{n-1}>$}{
        $v_p = null, i=0, voxline=<>$\;
        \While{$i < n$}{
            $v_i = \lfloor (p_i - B_{min}) / s \rfloor$ \;
            \If{$v_p \neq v_i$}{
                \If{$i \neq n - 1$}{
                    $voxline \gets voxline + p_{i}$ \;
                }
                \eIf{ $voxelset \textup{\textbf{ contains key }} v_p$}{
                $voxelset[v_p] \gets voxelset[v_p] \cup \{ voxline \} $ \;
                }{
                 $voxelset[v_p] \gets  \{ voxline \}$
            }
                $v_p = v_i, voxline = \emptyset$
            }
            $voxline \gets voxline + p_i$ \;
            $i = i + 1$
        }
        \If{$voxline \neq \emptyset$}{
            \eIf{ $voxelset \textup{\textbf{ contains key }} v_p$}{
                $voxelset[v_p] \gets voxelset[v_p] \cup \{ voxline \} $ \;
                }{
                 $voxelset[v_p] \gets  \{ voxline \}$
            }
        }
    }
    \KwResult{Set of voxlines grouped by voxel coordinates.}
    \caption{Voxline algorithm. Generates voxlines grouped by voxel coordinates from the input streamlines. }
\end{algorithm}
Algorithm \ref*{voxlines-algo} implements the voxelization \index{voxelization} and voxeline generation described in Sections \ref*{voxelization} and \ref{generating-voxlines}. It generates the set of voxlines and groups them based on voxel coordinates by iterating over the streamlines. Each streamline is split into voxlines whenever a point is found with a different voxel coordinate than the previous point. Voxlines that do not contain the final point have an additional point added to ensure proper connectivity between voxlines.

\subsection{Mesh generation} \label{mesh-generation}
We generate a mesh per voxel containing all voxlines part of the voxel, which can be rendered using the OpenGL \index{OpenGL} 'lines' primitive. A mesh consists of vertices, each with a position and indices that define the render order of the lines. The mesh vertices are simply defined by the points of each voxline that fall within the voxel. We define the indices by concatenating each pair of consecutive voxline points within a voxel. The order of these pairs in the indices determines the rendering order of the line segments within a voxel. For now, we define this order simply based on the dataset order. Later, we will describe a more sophisticated view-dependent order\index{view-dependent order} in Section \ref{view-dependent-line-orders}. Note that this method focuses solely on vertex positions and indices to determine the render order of the lines. Additional vertex information can be stored for cosmetic purposes, such as line identifiers, relative point indices, line tangents, and render flags. However, this additional data is not relevant for the render method as we are only concerned with vertex positions and the render order for improved transparency.

\subsection{Render Order Accuracy}
Transparency issues become most apparent when distant line segments are rendered in the incorrect order, specifically when line segments closer to the view are drawn before line segments further away. To address this, we divided the dataset into voxels, with each voxel representing the line segments passing through its bounds. Additional line segments were included to connect different voxel meshes. Prior to rendering the voxel, we sort them from back to front based on the current view. This sorting provides guarantees on the render order between any two line segments when the distance between any two consecutive streamline points is smaller than the voxel size. Line segments that are fully contained within different voxel bounds will be rendered in the correct back-to-front order. This holds true for the majority of line segments. Line segments that cross voxel bounds may have inaccurate render order relative to their immediate neighboring voxels. However, they are guaranteed to be in the correct order relative to line segments in non-neighboring voxels. Since we have not specifically ordered the lines within a voxel, the render order of line segments within a voxel will be inaccurate. Therefore, voxelization ensures that any two line segments have an accurate render order unless they are within the same voxel or one voxel apart for line segments crossing a voxel boundary, effectively resolving the most significant render order issues.

\subsection{Improved Transparency within Voxels}
The primary issue remaining with the render order, and consequently transparency, is the inaccurate rendering order within a voxel. We can take advantage of the fact that voxelization divides the dataset into smaller datasets with similar properties to the original streamline data. Since both the original streamline dataset and each voxel consist of sets of lines composed of consecutive points, we can apply existing rendering techniques that approximate transparency for 3D line sets on a voxel level. Furthermore, sorting within voxels becomes more feasible because each voxel contains a smaller subset of the points. For this method, we have developed a solution that stores different streamline orders per voxel to improve transparency.

\subsection{View-Dependent Line Order per Axis} \label{view-dependent-line-orders}
To achieve more accurate transparency within voxels, we can sort the line segments based on the average position of the two endpoints and the current viewing direction. However, performing this sorting process every time the view changes would be computationally expensive, even after voxelization. Instead, we precompute line render orders for a set of viewing directions. When rendering, we select the precomputed "closest" viewing direction for improved transparency. The closest viewing direction is determined by calculating the distance between each stored sorting direction and the current viewing direction using the formula $\boldsymbol{s \cdot \frac{v - c}{||v - c||}}$, where $\boldsymbol{s}$ represents the sorting direction, $\boldsymbol{v}$ is the voxel position, and $\boldsymbol{c}$ is the camera position.

This method is similar to imposter rendering, a technique that precomputes textures based on specific camera angles and then renders billboards instead of meshes for improved performance. However, instead of precomputing textures, we precompute line orders for a set of view directions. For the set of view directions, we use the six vectors defined by positive and negative unit vectors along each axis ($X$, $Y$, and $Z$). These six vectors are particularly well-suited as they are faster to sort than others, requiring no squared distance calculations. Additionally, these axis viewing directions align with axial, sagittal, and coronal projections, commonly used in tractography tools.

To sort the line segments for an arbitrary direction $\boldsymbol{d}$, we can compare $\boldsymbol{\frac{(p_0 + p_1)}{2} \cdot d}$ for each line segment, where $\boldsymbol{p_0}$ and $\boldsymbol{p_1}$ represent the two points making up the line segment. Since we only use unit vectors for $\boldsymbol{d}$, we can simplify the comparison to a single vector component, comparing the non-zero component of $\boldsymbol{d}$. Additionally, we optimize the sorting by using the sum of the two line positions instead of the average position. For example, sorting line segments for the direction $(0,1,0)$ would involve comparing the $Y$ component of $\boldsymbol{p_0 + p_1}$ for each line segment in the voxel. To find the orders of the negative unit vectors, we simply reverse the orders found for the positive unit vectors.

\subsection{Evaluation}
To evaluate our proposed approach, we generated a set of whole-brain tractograms from a single participant sourced from the Human Connectome Project \cite{HCP}. A tractogram consisting of one million streamlines was constructed using multi-shell multi-tissue constrained spherical deconvolution \index{spherical deconvolution} in MRtrix.

\begin{figure}[H]
    \includegraphics[width=\textwidth]{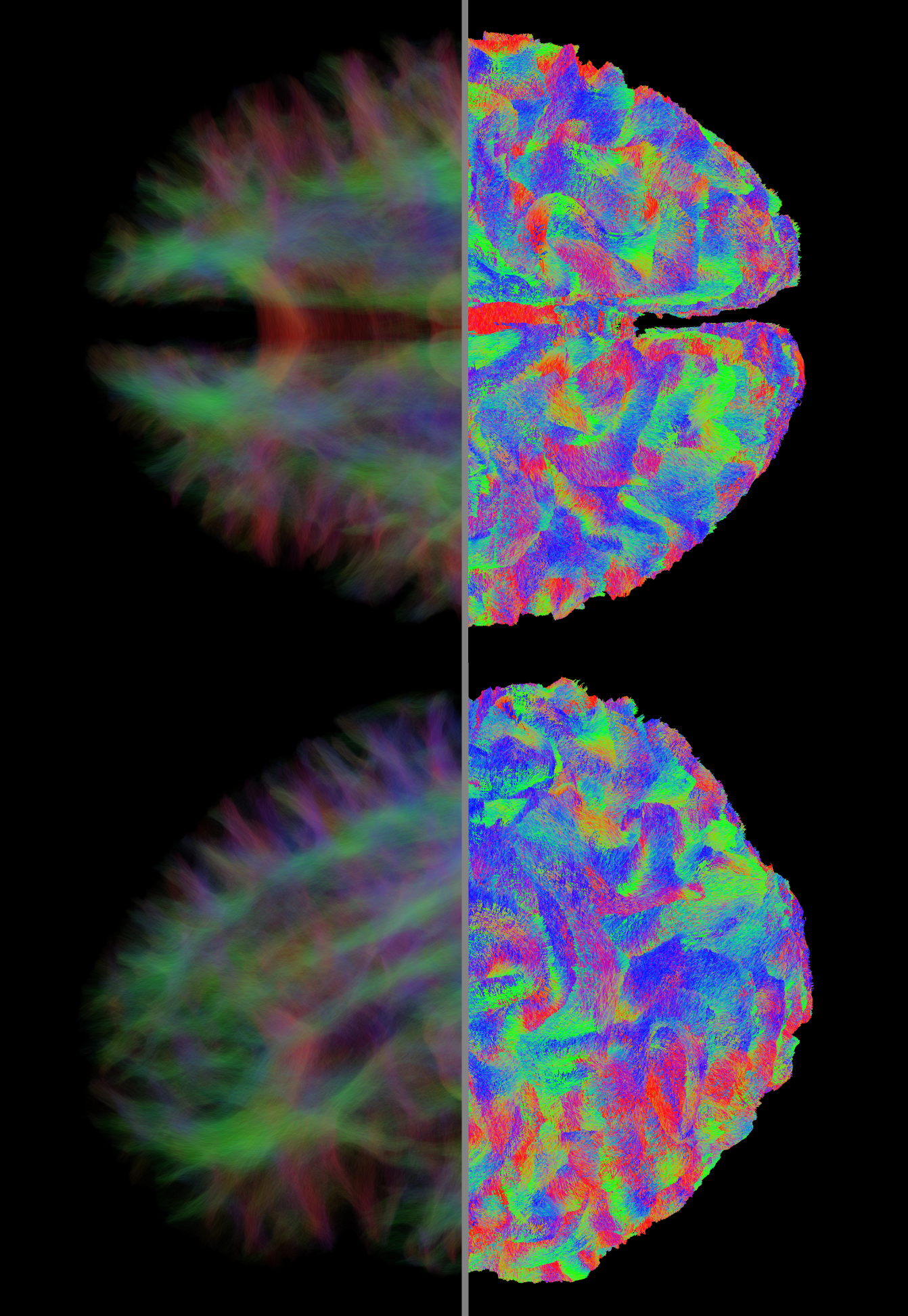}
    \caption{This figure shows two different angles of a dataset consisting of one million streamlines rendered using our method with voxelization and axis view-dependent line orders. Each image is divided into two halves, with the left half rendered with 0.5\% streamline opacity, and the right half rendered without transparency.}\label{fig4}
\end{figure}

Subsequently, a smaller dataset was created using TractSeg \index{TractSeg} \cite{TractSeg}, resulting in a tractogram with 140k streamlines. We implemented our render method in NeuroTrace (gitlab.com/Besm/NeuroTrace), a tractography visualization tool developed by the authors. We compared the visual and performance results of our method with two popular tractography visualization tools that support transparency, namely MRtrix and TrackVis. We compare them to our method with only voxelization (basic) and the extension with axis based view-dependent line orders (axis). All results use voxel size $10$mm$^3$, which through experimentation was found to give a balance between performance and visual quality for both individual bundles and whole brain tractograms.

\section{Results} \label{results}
\subsection{Qualitative}

\begin{figure}[H]
    \includegraphics[width=\textwidth]{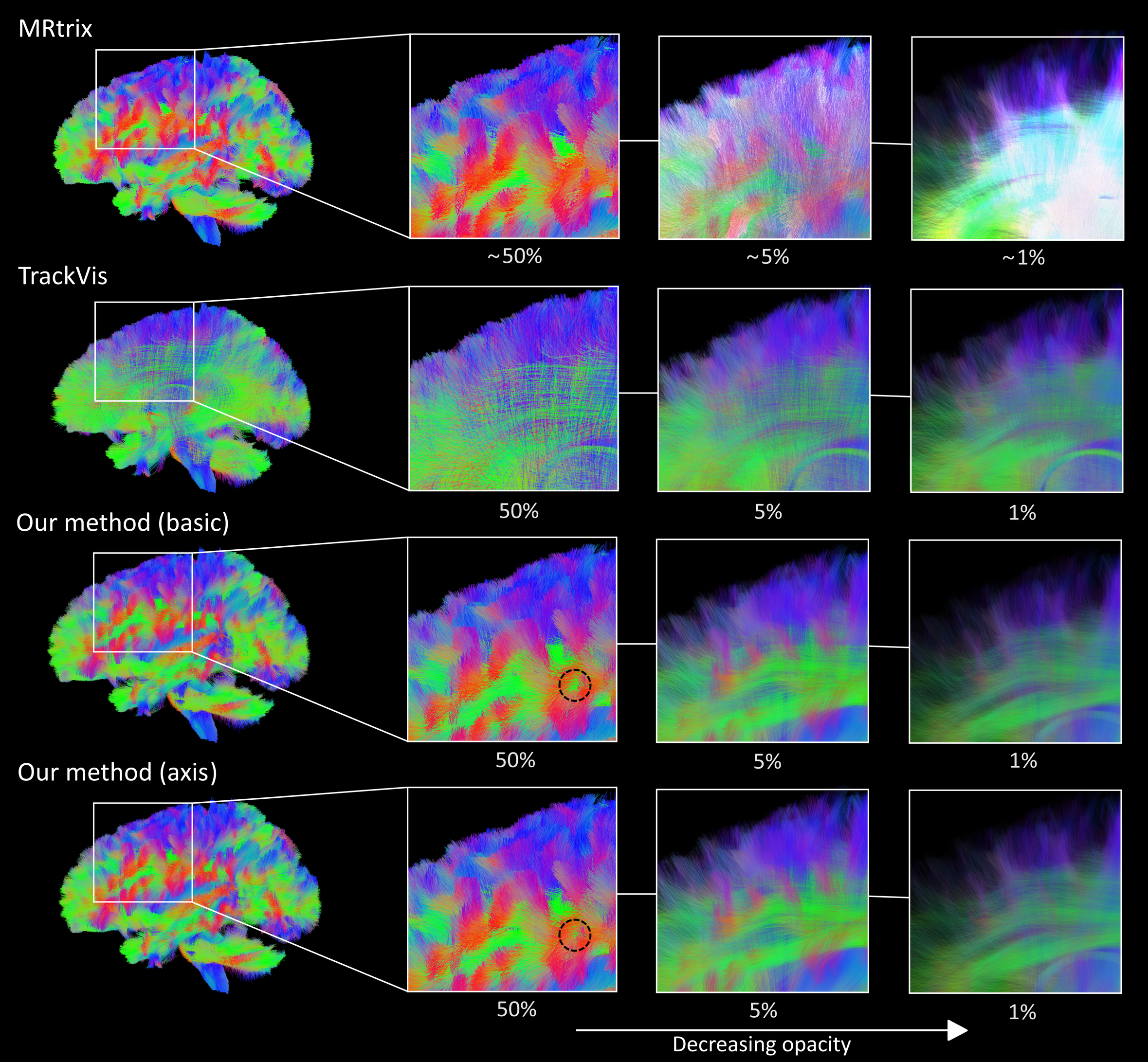}
    \caption{Figure 3 shows a comparison of transparency between MRTrix, TrackVis, and our method, with just voxelization (basic) and with view-dependent internal voxel order (axis). For each method, we display the entire dataset with $50\%$ transparency value from the sagittal plane and highlight a portion of the data with three opacity values: $50\%$, $5\%$, and $1\%$. Since MRTrix does not display opacity values, three comparable values were selected for the comparison.}\label{transparency-comparison}
\end{figure}

Figure \ref{transparency-comparison} displays different transparency values for each method using a single dataset. MRTrix handles transparency by blending two renders of the dataset,  one with a depth buffer and another without, based on the transparency value of the dataset \cite{MRtrix3-issue}. While this approach preserves line render order, it makes it challenging to visualize the internal structures of the dataset. TrackVis renders transparent streamlines based on the given dataset, causing noticeable render order issues, particularly when streamline transparency is low.

In our method, by utilizing only voxelization (basic), we can observe more information about the streamlines deeper in the brain with minimal sorting inaccuracy. This is evident in the figure, for example, green superior longitudinal fasciculus \index{superior longitudinal fasciculus} fibers can be seen behind the red, short superficial fibers \index{superficial fibers}. Increasing transparency increases the visibility of the superior longitudinal fasciculus fibers without distorting the fact that the short superficial fibers are in front, while MRTrix and TrackVis have more difficulty conveying this information.

When applying axis sorting, the render order issues are slightly reduced. Although it may be challenging to discern in the image, axis method gives a more accurate representation of the render order compared to basic method. For instance comparing highlighted regions in Figure \ref{transparency-comparison} shows axis sorting bringing out certain red superficial fibers in front more clearly compared to the method without axis sorting. 

\subsection{Quantitative}
Table \ref{performance-table} presents the performance comparison of our method with MRTrix and TrackVis on two tractography datasets: one with 140k streamlines and another with 1 million streamlines. It is worth noting that some tractography rendering methods include a preprocessing step that converts the dataset to a different format, aiming to reduce loading time for subsequent dataset loading. In our method, we intentionally omitted this preprocessing step to ensure a fair time comparison with MRTrix and TrackVis. Our implementation load the same TCK file format as MRTrix. 

\begin{table}[h]
    \centering
    \begin{tabular}{cccccc}
        \hline
        \multirow{2}{*} & \multirow{2}{*}{\textbf{Streamlines}} & \multirow{2}{*}{\textbf{Points}} & \multirow{2}{*}{\textbf{Method}} & \multirow{2}{*}{\textbf{Loading time (sec) }} & \multirow{2}{*}{\textbf{Render times (ms) }} \\
                        &                                       &                                  &                                  &                                               &                                              \\
        \hline
        \multirow{4}{*} & \multirow{4}{*}{143,999}              & \multirow{4}{*}{3,770,127}
                        & MRTrix                                & 0.8                              & 14                                                                                                                              \\
                        &                                       &                                  & TrackVis                         & 19.8 (+2)                                     & 30                                           \\
                        &                                       &                                  & our method (basic)               & 3.6                                           & 26                                           \\
                        &                                       &                                  & our method (axis)                & 6.5                                           & 34                                           \\
        \hline
        \multirow{4}{*} & \multirow{4}{*}{1,000,000}            & \multirow{4}{*}{56,240,953}
                        & MRTrix                                & 2.6                              & 67                                                                                                                              \\
                        &                                       &                                  & TrackVis                         & 283 (+26)                                     & 253                                          \\
                        &                                       &                                  & our method (basic)               & 27                                            & 124                                          \\
                        &                                       &                                  & our method (axis)                & 143                                           & 330                                          \\
        \hline                                                                                                                                                                                                                       \\
    \end{tabular}
    \caption{Performance comparison between different methods with transparency. Loading time is measured from selecting the dataset (TCK/TRK file) to first render of the dataset. Performance is measured on a Windows Laptop with NVIDIA GeForce GTX 1650 and Intel(R) Core(TM) i7-9750H 2.60GHz CPU. TrackVis has additional loading time whenever modifying transparency value, which is shown in parenthesis.}\label{performance-table}
\end{table}

\section{Discussion}
Voxelization and view-dependent line orders have shown promising results. The main benefits are from the voxelization step. View-dependent line order per voxel improve visual results at a computational cost, allowing for increased quality or larger voxel sizes with equivalent visual quality. Although we have not achieved the same performance as MRtrix, both our rendering and loading times are usable and faster than TrackVis. We believe that the decreased performance is caused by our implementation and is unrelated to the method outlined in this paper. In our implementation loading a dataset without voxelization has similar performance to loading with voxelization, indicating suboptimal performance in dataset loading independent of our method. The tractography visualization tool we used to implement this method, NeuroTrace, is primarily focused on functionality and experimentation rather than performance optimization. We aim to improve performance in future work.

Currently, we manually determine the voxel size and whether to generate the view-dependent render orders. We are exploring the possibility of utilizing metadata of the dataset, such as step size and total point count, to automatically determine the parameters of our method. We are experimenting with varying transparency per streamline to highlight specific parts of the dataset and utilizing the benefits gained from voxelization in novel ways. We aim to explore these aspects in future work.

\subsection{Different sorting orders}
\begin{figure}[t]
\includegraphics[width=\textwidth]{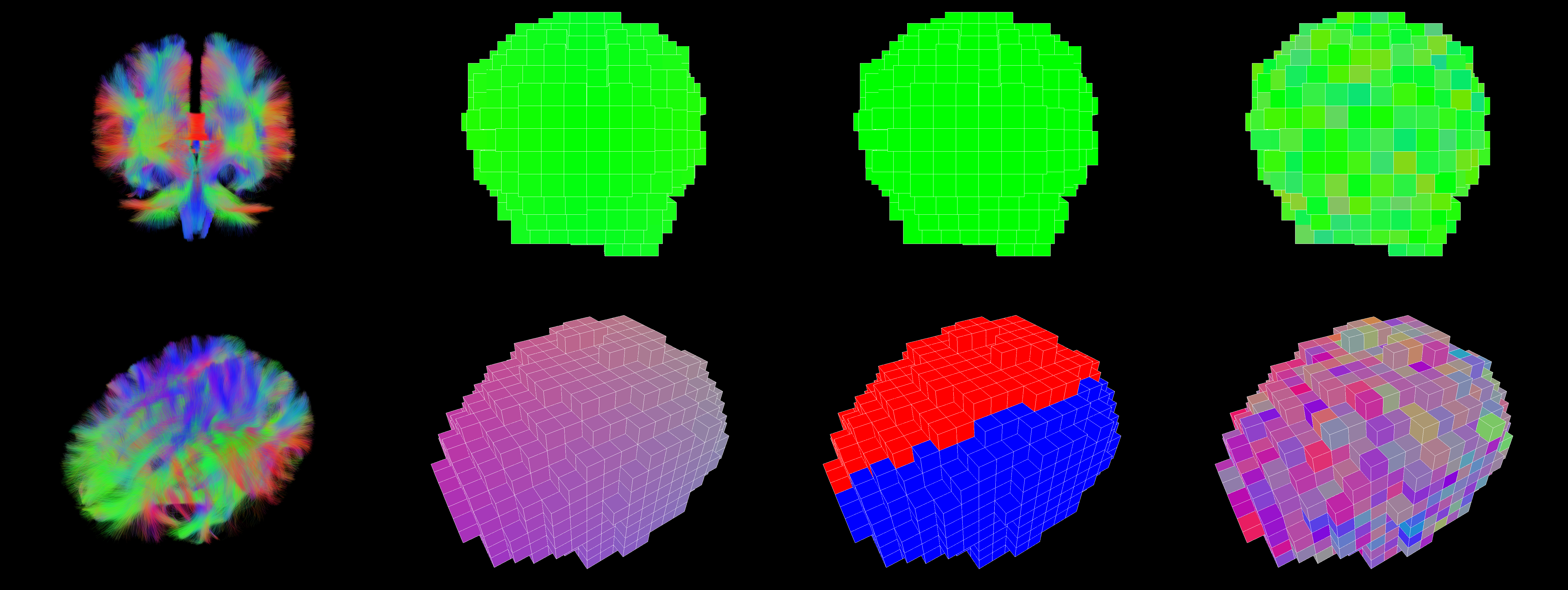}
\caption{This figure illustrates the voxel line ordering accuracy for two different camera angles. Each voxel is colored according to the absolute value of the direction for the RGB values. This direction differs per column. The second column displays the 'optimal sorting direction,' which is defined as the normalized direction from the camera to the voxel center. The third column shows the nearest voxel order direction using axis orderings. The fourth column shows the nearest voxel sorting direction using $64$ pseudo-random sorting directions.} \label{voxel-cubes-figure}
\end{figure}

Regarding the set of precomputed view direction line orders, we have investigated two approaches: per-axis sorting and pseudo-random sorting. The per-axis method used in this work involves generating two orders per axis. For each voxel, we store line render orders in both ascending and descending order for each axis. The pseudo-random sorting approach involves sorting points based on pseudo-random directions that differ for each voxel. Per-axis sorting offers certain benefits mentioned previously, mainly performance, as axis directions do not require (squared) distance calculations while sorting on any other direction does. Additionally, these axis viewing directions align with axial, sagittal, and coronal projections, which are common viewing angles in tractography tools. However, we encountered an issue with voxelization and per-axis method where moving the camera tends to update certain lines of voxels simultaneously, causing visual jittering when rotating the viewing direction while using high streamline opacity (besm.gitlab.io/voxlines/videos). Using the pseudo-random direction mitigates this issue but has less accurate results for the most common viewing direction and is more computationally expensive.

In Figure \ref{voxel-cubes-figure}, we demonstrate the difference between axis-based orders and pseudo-random direction orders. We can observe that the axis-based order is more accurate, when the dataset is viewed from one of the sides. Pseudo-random sorting performs worse in these cases. However, when considering arbitrary viewing directions, it outperforms the axis-based approach. Furthermore, when moving the camera, the pseudo-random method exhibits less visual jittering. However, the loading time for the pseudo-random method is worse. In future work, we aim to explore different sorting orders to find a balance between visual quality and performance.

\section*{Conclusion}
In this work, we have proposed and implemented a novel transparency method that improves tractography visualizations, allowing the capturing of deeper structures in tractograms. We demonstrate improved transparency and comparable performance to existing tractography tools. Our method achieves this by providing a novel approximate transparency, which enhances the visibility of structures in the deeper regions of the brain compared to existing methods.

\newpage
\bibliographystyle{splncs04}
\bibliography{paper}
\end{document}